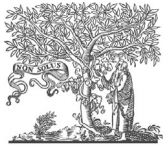
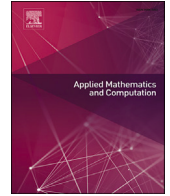

Full Length Article

# Time-dependent strategy for improving aortic blood flow simulations with boundary control and data assimilation

Muhammad Adnan Anwar [ID],*, Jorge Tiago [ID]

*Department of Mathematics and CEMAT - Center for Computational and Stochastic Mathematics, Instituto Superior Técnico, University de Lisboa, Av. Rovisco Pais, Lisbon, 1049-001, Portugal*

A B S T R A C T

Understanding time-dependent blood flow dynamics in arteries is crucial for diagnosing and treating cardiovascular diseases. However, accurately predicting time-varying flow patterns requires integrating observational data with computational models in a dynamic environment. This study investigates the application of data assimilation and boundary optimization techniques to improve the accuracy of time-dependent blood flow simulations. We propose an integrated approach that combines data assimilation methods with boundary optimization strategies tailored for time-dependent cases. Our method aims to minimize the disparity between model predictions and observed data over time, thereby enhancing the fidelity of time-dependent blood flow simulations. Using synthetic time-series observational data with added noise, we validate our approach by comparing its predictions with the known exact solution, computing the $L^2$-norm to demonstrate improved accuracy in time-dependent blood flow simulations. Our results indicate that the optimization process consistently aligns the optimized data with the exact data. In particular, velocity magnitudes showed reduced discrepancies compared to the noisy data, aligning more closely with the exact solutions. The analysis of pressure data revealed a remarkable correspondence between the optimized and exact pressure values, highlighting the potential of this methodology for accurate pressure estimation without any previous knowledge on this quantity. Furthermore, wall shear stress (WSS) analysis demonstrated the effectiveness of our optimization scheme in reducing noise and improving prediction of a relevant indicator determined at the postprocessing level. These findings suggest that our approach can significantly enhance the accuracy of blood flow simulations, ultimately contributing to better diagnostic and therapeutic strategies.

## 1. Introduction

Accurately modeling physiological processes, particularly blood flow, is crucial for understanding health and disease. Mathematical models, built upon physical principles, offer a quantitative framework to analyze complex biological systems [6,15,22]. These models, incorporating factors like blood vessel geometry, pressure gradients, and blood rheology, enable researchers to simulate blood flow patterns and predict potential disruptions. This understanding is vital for diagnosis, treatment planning, and drug development.

* Corresponding author.
 *E-mail addresses:* adnan.anwar@tecnico.ulisboa.pt (M.A. Anwar), jorge.tiago@tecnico.ulisboa.pt (J. Tiago).

Four-dimensional Magnetic Resonance Imaging (4D MRI) has become a powerful tool for non-invasively visualizing blood flow dynamics [18]. By capturing spatial and temporal information, 4D MRI allows researchers to observe blood flow velocities and track changes over time. This data provides valuable insights into the physiological function and helps identify potential abnormalities like stenosis (narrowing) or aneurysms (bulges) in blood vessels [7,25,28].

However, 4D MRI data often suffers from limitations. Noise and sparsity can hinder the accurate characterization of blood flow [4,14,16,24]. Here, Computational Fluid Dynamics (CFD) simulations come into play. CFD leverages mathematical models and numerical methods to simulate fluid flow in complex geometries, including blood vessels. By incorporating patient-specific data from 4D MRI or other sources, CFD simulations can provide detailed information on blood flow patterns and identify potential areas of concern.

Despite the advantages of CFD simulations, challenges remain. Accurate modeling requires complex computational resources and incorporates numerous parameters that may not be readily obtainable. Additionally, uncertainties in physiological data and limitations of the chosen model can lead to discrepancies between simulations and reality [2,3,5].

The so-called variational approaches, based on optimal control techniques, offer a promising approach to address these challenges. In fact, one may define an error functional to be minimized, while adhering to constraints in the form of a system of partial differential equations. Then, unknown boundary conditions, or eventually some physical or geometrical parameters, may be considered as control variables to be optimized, in order to minimize the error functional (see, for instance, [13,17,26]). In [8], the authors demonstrated, with stationary proof of concept examples, that stress type boundary controls, albeit being infinite dimensional variables, could be successfully used to reconstruct blood flow in some categories of arteries. Later, in [12], and [11] it was shown that the velocity profile itself could also be considered. In [10], the authors showed that the stationary assumption could also be dropped. This, however, came at the price of solving a reverse in-time adjoint equation for the application of a descent method, naturally increasing the total computational cost and losing some accuracy on the reconstruction. In [27], the authors dropped the infinite-dimensional assumption on the velocity boundary condition and assumed it to be parametrized with only 5 unknowns. It was emphasized that, even under this simplifying assumption, the computational cost can be very high. Additionally, some authors have noticed and analyzed the difficulties resulting from the presence of the noise and sparsity in the data, when using variational approaches, [9,19].

In this study, we devise a practical strategy to decrease the computational cost when using the velocity as an infinite dimension boundary control of the unsteady Navier-Stokes equations. We start by discretizing the state equation in time and then formulating the optimal control system accordingly. Additionally, we linearize the convection term and employ a one-shot method to the linear optimality system to obtain the solution at each time step. This modification leads to the resolution of a stationary problem at each time step, effectively reducing the computational complexity compared to traditional methods that require solving both forward and backward time-dependent problems. This approach was inspired in a similar approach used for the estimation of an elastic parameter done in [20]. Up to our knowledge it has not been attempted in an infinite dimensional setting. Subsequently, we conduct a comparative analysis on the reconstructed velocity, pressure, but also on some post-processed quantities such as flow rate and wall shear stresses.

The structure of this paper is as follows. In Section 2, we discuss the problem formulation and cost functional along with the proposed strategy, followed by the derivation of the weak form of the saddle point problem for the optimal system, which includes both the adjoint equations and the state equations. In Section 3, we focus on the discretization of the problem using the finite element method and details the solution technique employed in this study. We also cover the process of generating synthetic data with characteristics similar to those of 4D MRI in Section 4. In Section 5, we analyze and compare the computed solution with the exact CFD solution. In this section we discuss in detail about velocity field, pressure, volume flow and wall shear stress. Finally, in Section 6, we conclude with a discussion of the findings and their implications.

## 2. Problem formulation

This section explores a control problem, focusing on the state system and the associated minimization functional. The state system is defined by the time-dependent Navier-Stokes equations, which describe the dynamics of unsteady, incompressible, and viscous fluids. Our study is centered on optimization problem involving quadratic functional related to the flow dynamics. This problem features multiple velocity controls applied to various segments of the aorta boundary.

As a concrete illustration, let us denote the spatial domain by $\Omega \subseteq \mathbb{R}^d$, where $d = 3$, as depicted in Fig. 1. This domain is occupied by the fluid, while $\partial\Omega$ denote the boundary of the domain $\Omega$, which is the union of the control boundary $\Gamma_c = \Gamma_{c1} \cup \Gamma_{c2} \cup \Gamma_{c3} \cup \Gamma_{c4}$, the wall boundary $\Gamma_w$, and outflow boundary denoted by $\Gamma_{out}$. Mathematically, we can express this as:

$$\partial\Omega = \Gamma_c \cup \Gamma_{out} \cup \Gamma_w$$

An ideal time-dependent optimal control problem states:
minimize:

$$J_T(v, u) = \frac{1}{2} \int_0^T \int_\Omega |v(t) - v_{int}(t)|^2 \, dx \, dt + \frac{\alpha}{2} \int_0^T \int_{\Gamma_c} |\nabla u(t)|^2 \, d\Gamma_c \, dt \qquad (1)$$

where $\alpha \geq 0$, $u$ is a control variable at time $t$, and $v$ is the velocity field obtained as the weak solution of the time-dependent system described as follows:

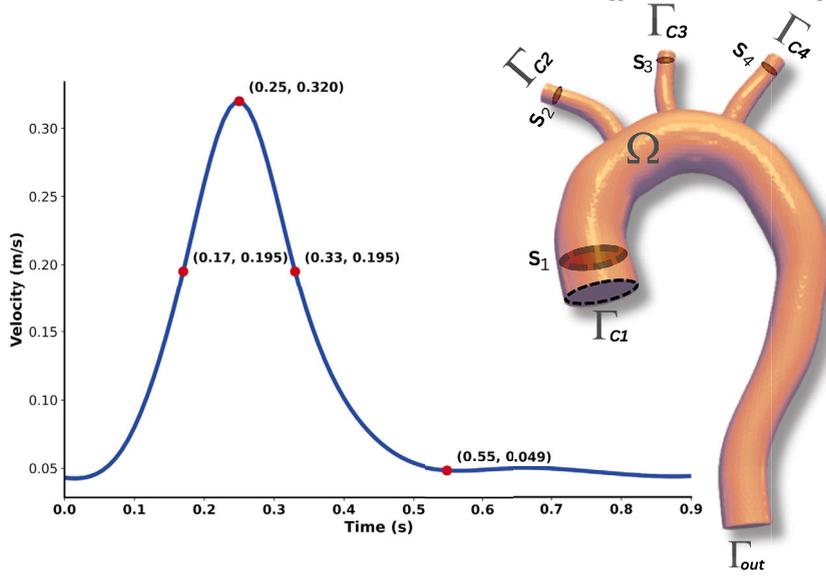

**Fig. 1.** Aorta geometry and inlet velocity profile with slices $S_1$, $S_2$, $S_3$ and $S_4$ near the control boundary.

$$\begin{cases} \dfrac{\partial v}{\partial t} - \mu \Delta v + (v \cdot \nabla)v + \nabla p = 0 & \text{in } \Omega \times (0,T] \\ \text{div } v = 0 & \text{in } \Omega \times (0,T] \\ v = 0 & \text{on } \Gamma_w \times (0,T] \\ v = u & \text{on } \Gamma_c \times (0,T] \\ \mu \dfrac{\partial v}{\partial n} - p\mathbf{n} = 0 & \text{on } \Gamma_{out} \times (0,T] \end{cases} \qquad (2)$$

where $p$ represents the pressure divided by dynamic viscosity and $\mu$ is the kinematic viscosity.

Let us first discretize the system (2) with respect to time. We adopt a classic implicit discretization of the linear terms (take $v = v_{n+1}, p = p_{n+1}, \kappa = 1/\Delta t$), together with a semi-discretization of the convective term. For more details about the convergence and stability properties of this approach see, for instance, [21].

This will transform the non-stationary problem into a sequence of stationary problems as follows:

$$\begin{cases} \kappa v - \mu \Delta v + (v_n \cdot \nabla)v + \nabla p = \kappa v_n & \text{in } \Omega \\ \text{div } v = 0 & \text{in } \Omega \\ v = 0 & \text{on } \Gamma_w \\ v = u & \text{on } \Gamma_c \\ \mu \dfrac{\partial v}{\partial n} - p\mathbf{n} = 0 & \text{on } \Gamma_{out} \end{cases} \qquad (3)$$

We can therefore consider the simplification of the cost functional for each time step as:

$$\min J(v,u) = \frac{1}{2}\|v(t_k) - v_{int}(t_k)\|^2_{L^2(\Omega)} + \frac{\alpha}{2}\|u(t_k)\|^2_{H^1_0(\Gamma_c)} \qquad (4)$$

and aim at minimizing (4) subject to (3).

### 2.1. Weak formulation

In order to write the weak formulation of system (3), assume that $u$ is a fixed function in $H^1_0(\Gamma_c)$, which means that $u$ vanishes on the boundary of $\Gamma_c$ and that its square, as well as the square of its tangential gradient, are both integrable functions. Therefore, $u$ also belongs to $H^{\frac{1}{2}}_{00}(\Gamma_c)$, the set which contains the continuous extensions (traces) of the velocity fields belonging to $H^1_{\Gamma_w}(\Omega) = \{v \in H^1(\Omega) \mid v|_{\Gamma_w} = 0\}$. We also introduce a new variable $s \in H^{-\frac{1}{2}}(\Gamma_c) = \left(H^{\frac{1}{2}}_{00}(\Gamma_c)\right)'$, accounting for the boundary stress at $\Gamma_c$ and the duality pair

$$\langle \cdot, \cdot \rangle_{\Gamma_c} = \langle \cdot, \cdot \rangle_{H^{-\frac{1}{2}}(\Gamma_c), H_{00}^{\frac{1}{2}}(\Gamma_c)}.$$

Consequently, we seek $(v, p, s) \in H^1_{\Gamma_w}(\Omega) \times L^2(\Omega) \times H^{-\frac{1}{2}}(\Gamma_c)$:

$$\begin{cases} a(v,\eta) + c(v_n; v,\eta) + b(p,\eta) + \langle s, \eta \rangle_{\Gamma_c} = \kappa \langle f, \eta \rangle, & \forall \eta \in H^1_{\Gamma_w}(\Omega), \\ b(v,\psi) = 0, & \forall \psi \in L^2(\Omega), \\ \langle \lambda, v \rangle_{\Gamma_c} = \langle \lambda, u \rangle_{\Gamma_c}, & \forall \lambda \in H^{-\frac{1}{2}}(\Gamma_c), \end{cases} \quad (5)$$

where:

$$a(v, \eta) = \kappa \int_\Omega v \cdot \eta \, d\Omega + \int_\Omega \mu \nabla v : \nabla \eta \, d\Omega, \quad c(v_n; v, \eta) = \int_\Omega (v_n \cdot \nabla) v \cdot \eta \, d\Omega,$$

$$b(\eta, p) = -\int_\Omega p(\nabla \cdot \eta) \, d\Omega, \quad \langle f, \eta \rangle = \int_\Omega v_n \cdot \eta \, d\Omega,$$

are bilinear forms, and $\lambda$ can be understood as a multiplier for the condition $v = u$. Also, observe that, under enough regularity of $v$ and $p$, s is given by $s = \mu \frac{\partial v}{\partial n} - p\vec{n}$.

Now, we define the Lagrangian of the problem,

$$\mathcal{L}(v,p,s,u,z,q,l) = J(v,u) + a(v,z) + c(v_n;v,z) + b(p,z) + \langle s, z \rangle_{\Gamma_c} - \kappa \langle f, z \rangle + b(p,q) + \langle l, v - u \rangle_{\Gamma_c} \quad (6)$$

where $(z, q, l) \in H^1_{\Gamma_w}(\Omega) \times L^2(\Omega) \times H^{-\frac{1}{2}}(\Gamma_c)$ can be understood as the Lagrange multiplier.

By differentiating (6) with respect to the adjoint variables and equating these derivatives to zero, (with $\tilde{\mathcal{L}}'_{(\cdot)} = \mathcal{L}_{(\cdot)}((\tilde{v}, \tilde{p}, \tilde{s}), \tilde{u}, (\tilde{z}, \tilde{q}, \tilde{l}))$), we obtain the following result:

$$\begin{cases} \tilde{\mathcal{L}}'_z \eta = a(\tilde{v},\eta) + c(v_n;\tilde{v},\eta) + b(\tilde{p},\eta) + \langle \tilde{s}, \eta \rangle_{\Gamma_c} - \kappa \langle f, \eta \rangle = 0, & \forall \eta \in H^1_{\Gamma_w}(\Omega), \\ \tilde{\mathcal{L}}'_q \psi = b(\tilde{v},\psi) = 0, & \forall \psi \in L^2(\Omega), \\ \tilde{\mathcal{L}}'_l \zeta = \langle \zeta, \tilde{v} \rangle_{\Gamma_c} - \langle \zeta, \tilde{u} \rangle_{\Gamma_c} = 0, & \forall \zeta \in H^{-\frac{1}{2}}(\Gamma_c). \end{cases} \quad (7)$$

We remark that system (7) corresponds to the weak form of the state problem given in system (5). Differentiating the Lagrangian's with respect to the $v, p, s$ and putting these derivatives equal to zero, we obtain:

$$\begin{cases} \tilde{\mathcal{L}}'_v \Theta = \langle \tilde{v} - v_{int}, \Theta \rangle + a(\Theta, \tilde{z}) + c(v_n; \Theta, \tilde{z}) + \langle \tilde{l}, \Theta \rangle_{\Gamma_c} + b(\Theta, \tilde{q}) = 0, & \forall \Theta \in H^1_{\Gamma_w}(\Omega), \\ \tilde{\mathcal{L}}'_p \Psi = b(\tilde{z}, \Psi) = 0, & \forall \Psi \in L^2(\Omega), \\ \tilde{\mathcal{L}}'_s \xi = \langle \tilde{z}, \xi \rangle_{\Gamma_c} = 0, & \forall \xi \in H^{-\frac{1}{2}}(\Gamma_c) \end{cases} \quad (8)$$

System (8) describes the weak form of the adjoint problem given by:

$$\begin{cases} \kappa \tilde{z} - (v_n \cdot \nabla) \tilde{z} - \mu \Delta \tilde{z} + \nabla \tilde{q} = \tilde{v} - v_{int} & \text{in } \Omega, \\ \text{div } \tilde{z} = 0 & \text{in } \Omega, \\ \tilde{z} = 0 & \text{on } \Gamma_c \cup \Gamma_w, \\ \tilde{q}\mathbf{n} - \mu \frac{\partial \tilde{z}}{\partial \mathbf{n}} - (v_n \cdot n) \cdot \tilde{z} = 0 & \text{on } \Gamma_{out}, \end{cases} \quad (9)$$

with $\tilde{l} = \tilde{q}\mathbf{n} - \mu \frac{\partial \tilde{z}}{\partial \mathbf{n}}$ on $\Gamma_c$. Ultimately, by computing the derivative of equation (6) with regards to $u$, we derive the following Euler equation:

$$\tilde{\mathcal{L}}'_u \lambda = \langle \alpha \tilde{u}, \lambda \rangle_{H^1_0(\Gamma_c)} - \langle \tilde{l}, \lambda \rangle_{\Gamma_c} = 0 \quad \forall \lambda \in U = H^1_0(\Gamma_c). \quad (10)$$

For an extended discussion on the well-posedness of system (7), (8), (10), we refer to [17] and the references therein.

## 3. Numerical space discretization

Let $X_h \subset H^1_{\Gamma_w}(\Omega)$ and $Q_h \subset L^2(\Omega)$ be finite element approximation subspaces of dimensions $N_V$ and $N_Q$ respectively.

The pressure and velocity fields' finite element approximations are $p_h \in Q_h$ and $v_h \in X_h$. The finite element approximations for the velocity and pressure fields of adjoint system are $z_h \in X_h$ and $q_h \in Q_h$. The control variable is approximated as $u_h \in U_h \subset U = H^1_0(\Gamma_c)$.

**Table 1**
Mesh convergence analysis with different resolutions of mesh.

| Mesh | Nodes | Degrees of Freedom (DOF) | $\|\|v_{\text{level}_i} - v_{\text{level}_5}\|\|_{L^2}$ | $\|\|z_{\text{level}_i} - z_{\text{level}_5}\|\|_{L^2}$ | $\|\|u_{\text{level}_i} - u_{\text{level}_5}\|\|_{L^2}$ |
|---|---|---|---|---|---|
| Level 1 | 6838  | 296918 | $4.44 \times 10^{-2}$ | $3.99 \times 10^{-4}$ | $1.18 \times 10^{-5}$ |
| Level 2 | 10609 | 472178 | $2.67 \times 10^{-2}$ | $2.49 \times 10^{-4}$ | $9.46 \times 10^{-6}$ |
| Level 3 | 13906 | 626516 | $1.93 \times 10^{-2}$ | $1.73 \times 10^{-4}$ | $8.28 \times 10^{-6}$ |
| Level 4 | 16312 | 740282 | $1.62 \times 10^{-2}$ | $1.46 \times 10^{-4}$ | $6.46 \times 10^{-6}$ |
| Level 5 | 19197 | 875994 | 0.0 | 0.0 | 0.0 |

Hence, the discretization of the coupled problem (7), (8), (10) reads as: find

$$(v_h, p_h, s_h, u_h, l_h, z_h, q_h) \in (X_h, Q_h, U_h, U_h, U_h, X_h, Q_h),$$

such that

$$\mathbb{A}\mathbb{U} = \mathbb{F}, \tag{11}$$

where $\mathbb{U}$ is the vector gathering the coefficients of the linear combinations defining $(v_h, p_h, s_h, u_h, l_h, z_h, q_h)$. We refer to [17] for more details on the well-posedness of the saddle point problem (11).

Since our problem is a large scale linear system, we need to employ an advanced linear solver to compute $\mathbb{U}$. We choose to use the built-in linear solver MUMPS, which we described in 3.2.

### 3.1. Numerical parameters

To obtain the discrete optimality system (11) we fixed $X_h = U_h = P_2$ for the velocity type variables and $Q_h = P_1$ for the pressure type variables. A convergence analysis for the solution $\mathbb{U}$, with respect to mesh refinement, was done and the results can be seen in Table 1, for optimal velocity $v_h$, adjoint velocity $z_h$ and control variable $u_h$.

From these results, we chose to perform our data assimilation using the level 3 mesh, which corresponds to 626516 degrees of freedom for the optimality system. Besides, to reconstruct the time-dependent velocity and pressure fields, we used a fixed time-step of 0.009, which suffices for the reconstruction of the sparse -in time- data, as we will see in the section 4. These choices allowed for a balance between computational efficiency and solution accuracy.

### 3.2. Solving the sparse linear system using MUMPS

To solve the complex sparse linear system $\mathbb{A}\mathbb{U} = \mathbb{F}$, we employed the Multifrontal Massively Parallel Sparse Direct Solver (MUMPS), for more detail, see [1]. MUMPS utilizes a multifrontal method to factorize the sparse matrix $\mathbb{A}$ into lower and upper triangular matrices, $\mathbb{L}$ and $\mathbb{U}$, while leveraging permutation matrices $\mathbb{P}$ and $\mathbb{Q}$ for enhanced stability and sparsity preservation. This factorization can be expressed as:

$$\mathbb{A} = \mathbb{PLUQ}.$$

The multifrontal method constructs and factorizes smaller dense submatrices called frontal matrices, efficiently managing fill-in and allowing for parallel execution. MUMPS employs dynamic pivoting strategies to ensure numerical stability, particularly for indefinite and symmetric matrices, represented by:

$$\mathbb{A} = \mathbb{PLDUQ},$$

where $\mathbb{D}$ is a diagonal scaling matrix.

The solution process involves forward and backward substitution:

- **Forward Substitution**: Solve $\mathbb{L}\mathbf{y} = \mathbb{P}\mathbb{F}$ for $\mathbf{y}$.
- **Backward Substitution**: Solve $\mathbb{U}\mathbf{x} = \mathbf{y}$ for $\mathbf{x}$.

In summary, MUMPS efficiently handles large-scale sparse systems by minimizing fill-in and exploiting parallelism, significantly enhancing computational performance and accuracy.

## 4. Synthetic data

To evaluate the efficacy of our approach, we utilized synthetic data designed to closely mimic the original data acquired from 4D MRI. This synthetic dataset is generated by solving the Navier-Stokes equations with $P2/P1$ element on a coarse spatial mesh with level 2 (see Table 1) and coarse temporal mesh, with step size $\Delta t = 0.9/20 = 0.045$, (see Fig. 2). We considered pulsating inlet flow (see Fig. 1) and do-nothing conditions applied at the outlet $\Gamma_{out}$ and at the three branch exits $\Gamma_{c_2}, \Gamma_{c_3}, \Gamma_{c_4}$.

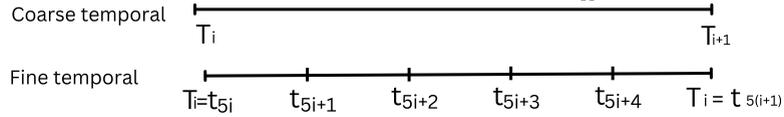

**Fig. 2.** Observed coarse temporal solution available only at $T_i$, for $i = 0, 1, 2, \cdots, 19$. In fine temporal, we added 4 time steps between any two given solution.

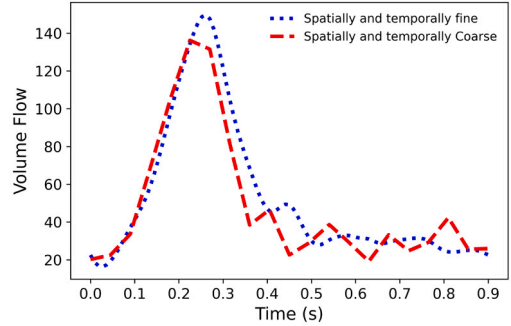

**Fig. 3.** Volume flow through aortic arch with the coarse and fine data set.

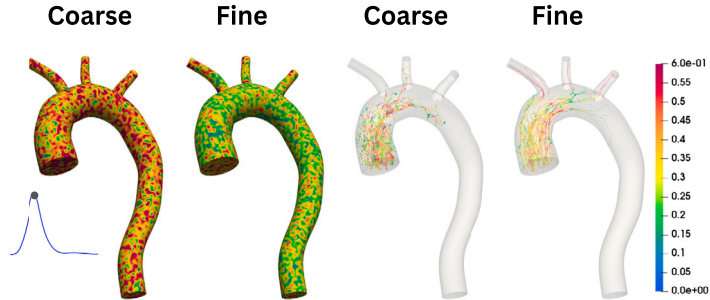

**Fig. 4.** These are the velocity magnitude and streamlines from inlet with coarse (without spatial and temporal interpolation) and fine (with spatial and temporal interpolation) data set.

Subsequent to solving the Navier-Stokes equations, Gaussian noise introduced to simulate measurement inaccuracies inherent in data acquired from 4D MRI. The noise was applied with a random temporal behavior, ensuring variability such that some time steps exhibited higher noise levels while others had lower noise levels. This stochastic approach to noise addition aimed to create a more realistic synthetic dataset.

The noise term can be represented mathematically as:

$$\text{noise} = \mathcal{N}\left(0, \frac{c \cdot u_0}{3}\right)$$

which denotes a normal (Gaussian) distribution with mean 0 and standard deviation $\frac{c \cdot u_0}{3}$, $c$ is a scaling factor, $u_0$ is maximum velocity at that time step.

To enhance spatial resolution, a spatial mesh at level 3 is selected, and temporal resolution is improved by interpolating four additional time steps between each original point, resulting in a reduced temporal step size of $\Delta t = 0.009$. This spatial and temporal discretization is chosen for simplicity in computations, while finer meshes can be employed for greater precision if desired.

Finally, the temporal interpolated solution mapped onto a finer spatial mesh through spatial interpolation. This step ensured that the synthetic data had the requisite spatial resolution for subsequent use in the optimal control problem. The resulting synthetic dataset, characterized by both high temporal and spatial resolution, served as the input for testing our optimal control framework, thereby enabling a thorough assessment of its performance in a controlled yet realistic setting.

Fig. 3 compares the volume flow rate through the aortic arch using both temporally and spatially coarse meshes with those obtained using temporally and spatially fine meshes. Fig. 4 presents the velocity magnitude on the aortic surface along with streamlines originating from the inlet, for both coarse and fine meshes in terms of spatial and temporal resolution. It is evident from Fig. 4 that the coarse mesh introduces more noise in the velocity and streamlines, whereas the use of spatial and temporal interpolation reduces this noise to some extent.

This rigorous synthetic data generation process ensured that our testing framework closely approximated real-world scenarios, thereby providing a meaningful platform for validating the proposed optimal control approach.

**Table 2**
Cumulative relative error (%) of velocity magnitude calculated with the formula given in (13).

| Avg. SNR value | Cumulative Relative Error | % Error |
| --- | --- | --- |
| 0.88 | 0.066 | 6.6 |
| 1.72 | 0.037 | 3.7 |
| 2.71 | 0.026 | 2.6 |

## 5. Result and discussion

In this section, we aim to present various results, including velocity, pressure, volumetric flow rate, and wall shear stress, to compare the optimized solution with the exact solution.

### 5.1. Velocity

Velocity plays a crucial role in understanding blood flow dynamics, as it directly impacts shear stress on vessel walls and overall circulatory efficiency. Accurate velocity measurements are essential for diagnosing and managing cardiovascular conditions, making the optimization of noisy data vital for clinical and research applications.

In the context of our optimal control problem, we have analyzed the velocity magnitudes at specific time steps, leveraging varying Signal-to-Noise Ratio (SNR). The visualization provided illustrates the comparative analysis across noisy input synthetic data, optimized solutions, and exact solutions, enabling a comprehensive assessment of the optimization effectiveness. The slice plots show the velocity magnitude within the aorta under different control conditions, enabling a detailed evaluation of the model's performance across various cardiac time points.

The slice plots in Fig. 5 show how the velocity magnitudes change over time across various SNR values, comparing the optimized, exact and interpolated solutions. These visualizations are crucial for understanding the flow dynamics within the aortic geometry under different simulation conditions. The "Noisy" plots indicate the presence of noise in the data, which likely simulates real-world conditions where measurements are subject to various levels of interference. Conversely, the "Optimized" plots aim to show the improvements gained through computational optimization techniques.

From the comparative analysis, it is evident that the optimized model significantly reduces the discrepancies observed in the noisy data. For instance, specific data points in the "Noisy" scenario show higher deviations from the "Exact" values, while the "Optimized" data points align more closely with the exact solutions. This indicates that the optimization approach effectively mitigates the noise, enhancing the accuracy of the velocity measurements. Such improvements are critical for applications requiring precise flow measurements and control, such as in medical diagnostics and surgical planning.

Furthermore, the observed data in the "Optimized" scenario consistently exhibits lower error margins compared to the noisy data, reinforcing the effectiveness of the optimization process. This detailed comparison underscores the importance of advanced computational methods in refining model predictions, ultimately leading to better-informed decisions in the management and treatment of aortic conditions. The technical evaluation of these slice plots thus provides a robust framework for assessing the impact of noise and optimization on the accuracy of computational fluid dynamics models in biomedical applications.

To better quantify the observed discrepancies, we will analyse the relative error which is computed using the formula:

$$E_{\text{rel}} = \frac{\|v - u_d\|_{L^2(\Omega)}}{\|u_d\|_{L^2(\Omega)}}, \quad (12)$$

which quantifies the deviation of the optimized velocity field from the exact solution $u_d$. The Fig. 6 presents the relative error (%) for various values of SNR at different time steps, corresponding to three different values of the parameter $c$ (1, 0.5, and 0.3). As the parameter $c$ decreases, the SNR generally increases, while the relative error decreases. The SNR ranges from 1.23 to 3.62, and the corresponding errors vary between 5.7% and 2.4%, depending on the time step and the value of $c$. The associated plots above the table in Fig. 6 illustrate the inlet velocity curve, highlighting a specific point within the cardiac cycle for reference in the error analysis.

The cumulative relative error is computed using the formula:

$$\text{Cumulative Relative Error} = \frac{\|v - u_d\|_{L^2(\Omega \times (0,T])}}{\|u_d\|_{L^2(\Omega \times (0,T])}}. \quad (13)$$

The Table 2 shows the cumulative relative error for different average SNR values. The SNR indicates the level of noise in the data, with higher values corresponding to cleaner data.

As the SNR value increases, indicating higher quality data with less noise, the cumulative relative error decreases. This trend illustrates the importance of high-quality data in achieving accurate computational solutions. For instance, with an average SNR of 0.88, the cumulative relative error is 0.066, which corresponds to a 6.6% error. When the average SNR improves to 2.71, the cumulative relative error drops to 0.026, resulting in a significantly lower error of 2.6%.

Funke et al. conducted a study focusing on time-dependent problems where they utilized a backward adjoint method in time. Their work involved optimization in a low-dimensional parameter space and reported relative errors with SNR of 2 ranging from 8% to 22% depending on the chosen time interpolation method in a 2D setting, see [10]. Our study extends this by addressing infinite-dimensional optimization in a 3D context. Specifically, we have achieved a significant improvement in error metrics, obtaining a relative error of 3.7% for an SNR of 1.72. This demonstrates the enhanced efficiency and accuracy of our approach in more complex and higher-dimensional scenarios.

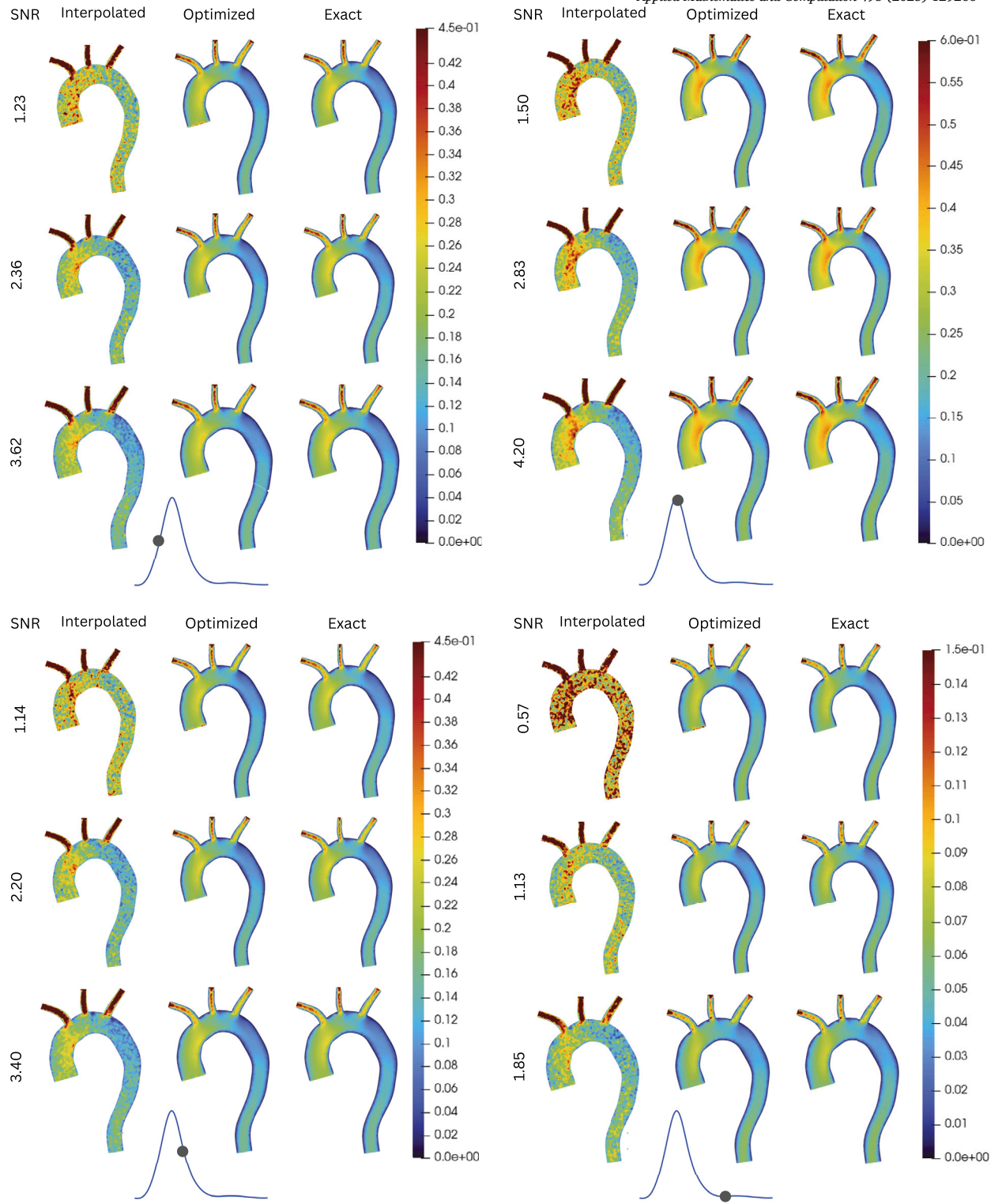

**Fig. 5.** Slice showing the velocity magnitude of 3D aorta against different time and SNR value.

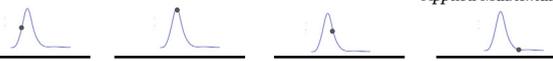

| c | SNR | % Error | SNR | % Error | SNR | % Error | SNR | % Error |
|---|---|---|---|---|---|---|---|---|
| 1 | 1.23 | 5.7 | 1.5 | 5.1 | 1.14 | 6.8 | 0.57 | 12 |
| 0.5 | 2.36 | 3.2 | 2.83 | 3.1 | 2.2 | 3.4 | 1.13 | 6.4 |
| 0.3 | 3.62 | 2.4 | 4.2 | 2.4 | 3.4 | 2.6 | 1.85 | 4.2 |

**Fig. 6.** Relative error for different values of SNR at different time step calculated with the formula given in (12).

This analysis emphasizes the effectiveness of the proposed methodology in delivering precise and reliable optimization results, particularly for biomedical applications where accuracy is crucial. The method significantly enhances the accuracy of simulations, reduces errors, and improves the robustness of model predictions, demonstrating its strong potential for generating actionable insights in complex biomedical scenarios.

### 5.2. Volume flow

Volume flow rate is crucial in cardiovascular research as it provides insights into the dynamics of blood circulation within arteries and veins, which is essential for diagnosing and treating various vascular diseases. Accurate measurement and analysis of volume flow rates can reveal abnormalities in blood flow, helping to detect conditions like stenosis or aneurysms early and allowing for effective intervention. Moreover, understanding volume flow is key to optimizing treatments and surgical procedures, ensuring better patient outcomes and advancing overall cardiovascular health.

The volume flow rate $Q(t)$ through a cross-sectional area $A$ is defined as:

$$Q(t) = \int_A v(x,t) \cdot \mathbf{n}\, dA, \tag{14}$$

where $v(x,t)$ is the velocity vector field as a function of position $x$ and time $t$, $\mathbf{n}$ is the unit normal vector to the cross-sectional area $A$, and $dA$ is the differential area element, where $v(x,t)$ is the velocity vector field as a function of position $x$ and time $t$, $\mathbf{n}$ is the unit normal vector to the cross-sectional area $A$, and $dA$ is the differential area element.

For numerical computations, the volume flow rate is approximated by summing the contributions from discretized surface elements at each time step:

$$Q(t) \approx \sum_{i=1}^{N} (v_i(t) \cdot \mathbf{n}_i)\Delta A_i, \tag{15}$$

where $N$ denotes the number of surface elements, $v_i(t)$ is the velocity vector at the $i$-th surface element at time $t$, $\mathbf{n}_i$ is the unit normal vector, and $\Delta A_i$ is the area of the $i$-th surface element.

The volume flow rate plots demonstrate the impact of noise and the effectiveness of the optimization approach across four distinct slices ($S_1$, $S_2$, $S_3$, and $S_4$), see Fig. 1.

In the Fig. 7, corresponding to an average SNR of 0.88, the noisy data exhibits significant fluctuations. For Slice $S_1$, there is a pronounced peak around 0.3 seconds, followed by a sharp decline, with the noisy data deviating considerably post-peak. The optimal data, though closely following the exact data initially, shows slight deviations after the peak, indicating effective noise reduction but with some remaining discrepancies. In Slice $S_2$, the peak is notably lower, with the noisy data showing high variability post-peak. The optimal data provides a smoother profile that aligns well with the exact data, indicating effective noise attenuation. For Slices $S_3$ and $S_4$, the volume flow rates display similar peaks, with the noisy data introducing considerable fluctuations. The optimal data closely follows the exact data but shows minor deviations, suggesting partial noise reduction.

In the Fig. 8, with an average SNR of 1.72, the noise levels are reduced. At Slice $S_1$, the noisy data still shows variability, but the optimal data aligns more closely with the exact data, indicating better noise reduction. At Slice $S_2$, the peak volume flow rate is lower, with the noisy data showing less variability compared to the previous SNR. The optimal data again closely follows the exact data, indicating improved noise attenuation. For Slices $S_3$ and $S_4$, the noisy data shows reduced fluctuations compared to the higher noise scenario. The optimal data follows the exact data well, suggesting effective noise reduction.

In the Fig. 9, with an average SNR of 2.71, compared to a scenario with higher noise, the data shows significantly reduced variability, particularly in the noisy plots, indicating successful noise reduction. Encouragingly, the "Optimal" data closely tracks the "Exact" data across all slices, suggesting the optimal control method effectively removes noise while preserving the underlying flow rate patterns. Interestingly, Slice $S_2$ exhibits a lower peak flow rate with less noisy data fluctuation, but the "Optimal" data still closely aligns with the "Exact" data. This reinforces the method's effectiveness across varying flow rate characteristics. Finally, while $S_3$ and $S_4$'s noisy data shows minimal fluctuations, the "Optimal" data in both slices still closely follows the "Exact" data, implying the noise reduction technique was beneficial even in inherently less noisy scenarios.

The analysis reveals a consistent pattern where the optimization process effectively reduces noise, as evidenced by the closer alignment of the optimal data with the exact data across all slices and SNR values. Higher SNR values correspond to cleaner data, with less variability in the noisy data and better performance of the optimal solution.

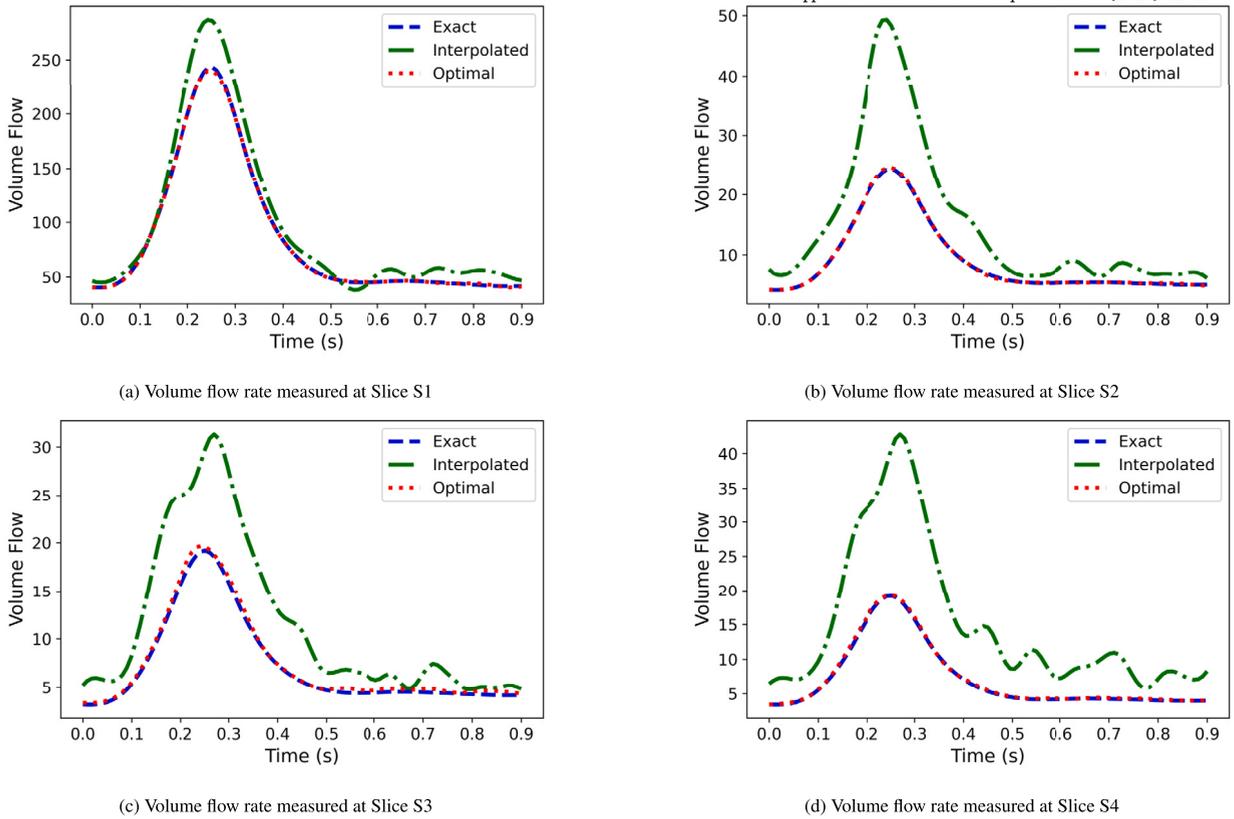

**Fig. 7.** Volume flow at c = 1 with average SNR 0.88.

**Table 3**
Relative error (%) of volume flow computed with the formula $\frac{\|exact-optimized\|_{L^2(\cdot,(0,T])}}{\|exact\|_{L^2(\cdot,(0,T])}} \times 100$.

| C | Avg SNR | Volume flow error (%) | | | | Pressure error (%) |
|---|---|---|---|---|---|---|
| | | $S_1$ | $S_2$ | $S_3$ | $S_4$ | aorta |
| 1 | 0.88 | 0.85 | 1.23 | 4.16 | 1.97 | 3.92 |
| 0.5 | 1.72 | 0.51 | 0.89 | 2.66 | 0.95 | 3.59 |
| 0.3 | 2.71 | 0.36 | 0.42 | 1.38 | 0.56 | 2.35 |

The results highlight the challenges posed by noise and the critical role of computational methods in enhancing the fidelity of biomedical simulations. The effectiveness of the proposed approach in reducing noise and improving the accuracy of volume flow rate measurements is evident from the comparisons across different SNR values.

Table 3 presents the relative errors (%) of volume flow and pressure computed at different SNRs demonstrating that, as the average SNR improves (indicating better signal quality and less noise), both the volume flow and pressure errors decrease significantly. This suggests that the optimization technique employed is effective in improving the accuracy of volume flow and pressure measurements in the aorta. The highest accuracy is observed at the lowest C value 0.3 with the highest SNR 2.71, confirming the robustness of the proposed method in reducing errors and enhancing simulation fidelity.

### 5.3. Pressure

In the Fig. 10, we have plotted the computed pressure together with the exact pressure. When direct data on pressure is unavailable, recovering pressure data from noisy velocity measurements becomes critical. Utilizing 4D MRI-type synthetic data, we have successfully extracted pressure values that show a remarkable correspondence with the exact pressure data. This recovery process involved interpolation and control techniques to filter out noise and accurately infer the pressure, ensuring that the computed data aligns closely with the exact values. The comparison, illustrated in the provided plots, demonstrates the efficacy of the method, where the computed average pressure on the whole aorta, and the computed pressure near the inlet (at a random node) nearly match the exact values throughout the time interval considered.

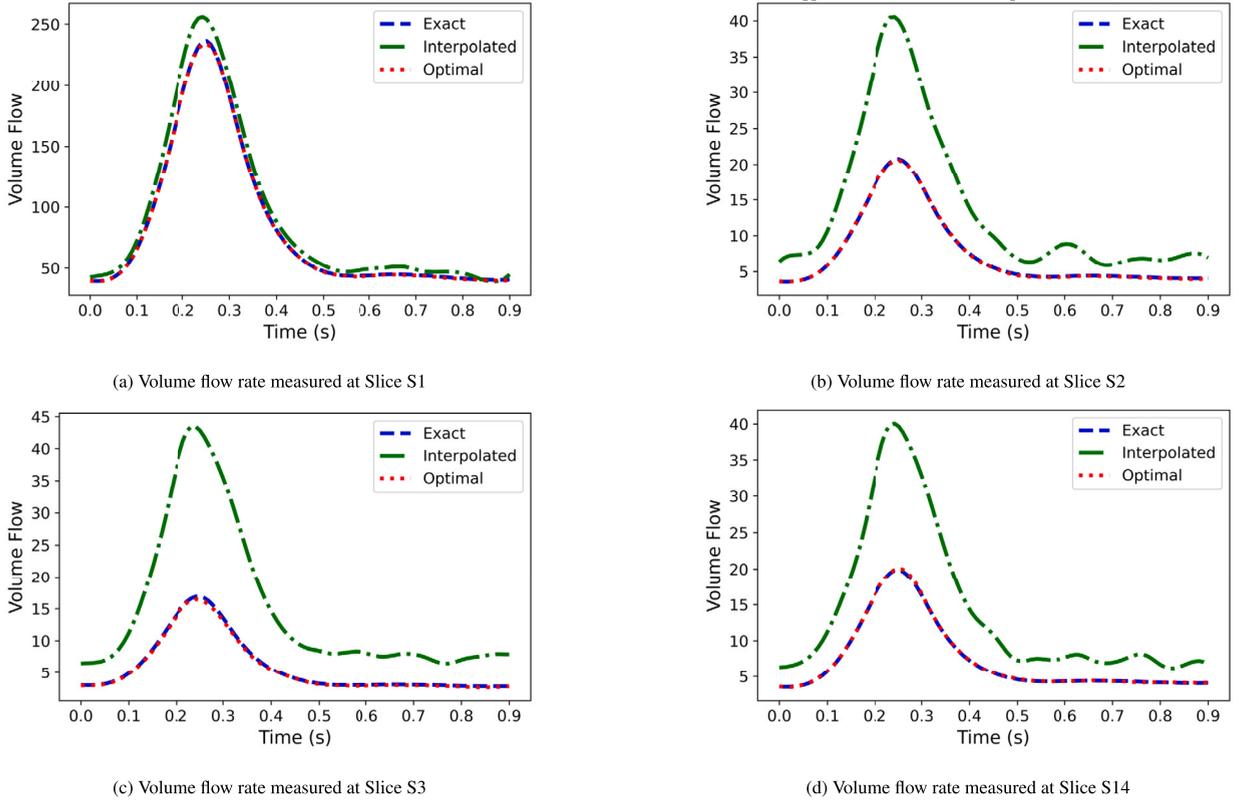

**Fig. 8.** Volume flow at c = 0.5 with average SNR ratio 1.72.

These findings are significant as they validate the approach of using synthetic 4D MRI-type velocity data to infer pressure in the absence of direct measurements. The close match between the recovered and exact pressures indicates the potential of this methodology for accurate pressure estimation, which is essential for solving optimal control problems in cardiovascular studies. This alignment not only enhances the reliability of the recovered data but also provides a robust basis for further analyses and simulations in medical research, where accurate pressure data is often critical yet challenging to obtain directly.

*5.4. Wall shear stress*

Wall shear stress (WSS) in the aorta is crucial as it influences endothelial cell function and vascular remodeling, impacting conditions such as atherosclerosis and aneurysms. For Newtonian fluids, the vectorial WSS is defined as:

$$\text{WSS} = (\mathbf{s} \cdot \mathbf{n}) - ((\mathbf{s} \cdot \mathbf{n}) \cdot \mathbf{n})\mathbf{n},$$

where $\mathbf{s}$ is the viscous stress vector [23], given by:

$$\mathbf{s} = \mu \left( \nabla \mathbf{u} + (\nabla \mathbf{u})^T \right) \mathbf{n},$$

where $\mu$ is the dynamic viscosity, $\nabla \mathbf{u}$ is the velocity gradient tensor, and $\mathbf{n}$ is the unit normal vector to the surface.

The Fig. 11 presents four sets of 3D surface plots, each corresponding to different time steps, showcasing the wall shear stress distribution in an aorta. In each set, there are three surface plots: one showing the WSS after applying the proposed optimization scheme (left), another displaying the exact WSS (center) and the third depicting WSS for interpolated data (right). The interpolated data plots exhibit significant irregularities and jagged variations, indicating the presence of substantial noise in the measurements, even after interpolation. In contrast, the optimized data plots demonstrate a significant reduction in noise, resulting in much smoother surfaces that closely match the exact WSS distributions. The exact WSS plots serve as a reference, displaying accurate and smooth surface patterns expected from high-fidelity simulations or precise measurements.

A curve below the surface plots illustrates the pulse time, indicating the specific moments at which each set of plots was captured, thereby providing a temporal context for the WSS variations. This comparison across noisy, optimized, and exact datasets highlights the effectiveness of the proposed optimization scheme. The optimized WSS surfaces consistently align more closely with the exact data than the noisy surfaces do, underscoring the scheme's ability to mitigate noise and enhance the accuracy of WSS predictions. This visual representation emphasizes the value of the optimization approach in improving the reliability of simulation results for biomedical and engineering applications.

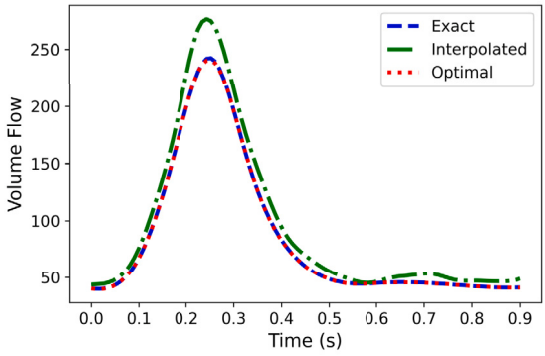

(a) Volume flow rate measured at Slice S1

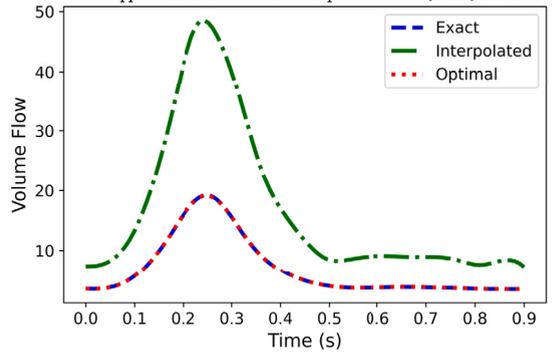

(b) Volume flow rate measured at Slice S2

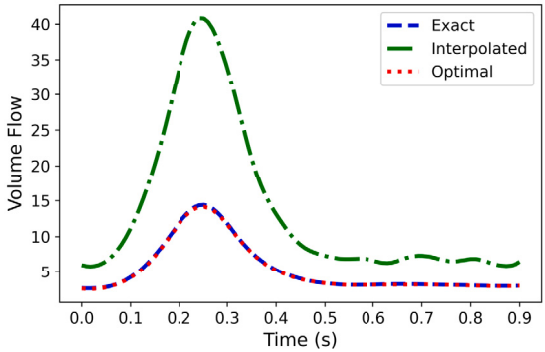

(c) Volume flow rate measured at Slice S3

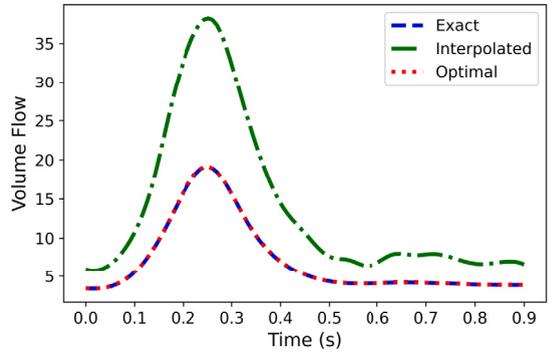

(d) Volume flow rate measured at Slice S4

**Fig. 9.** Volume flow at c = 0.3 with average SNR 2.71.

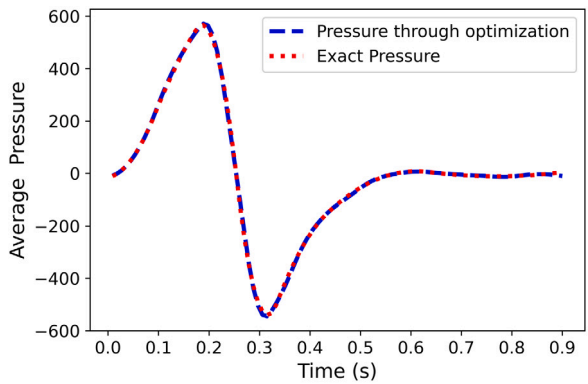

(a) Average pressure on the whole aorta along the time.

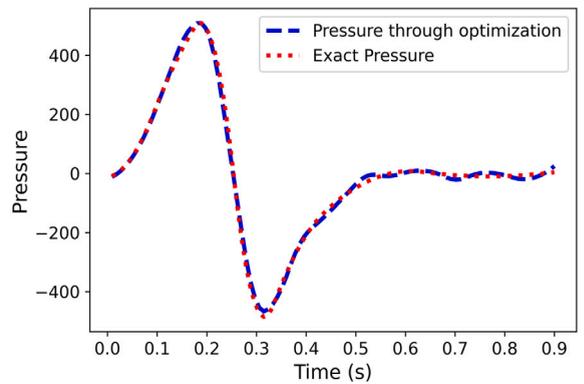

(b) Pressure near the inlet $\Gamma_{c1}$ at a random node.

**Fig. 10.** Pressure plot from the optimized solution without having any information about pressure computed from the 4D MRI type velocity data.

## 6. Summary and conclusions

In this study, we introduced an approach for addressing time-dependent, infinite-dimensional, optimal control problems (OCP) within the context of unsteady Navier-Stokes equations. Our methodology hinges on reformulating the problem, incorporating Dirichlet boundary control defined across multiple boundaries. This strategy involves discretizing the state equation in time, which allows us to tackle the OCP on a step-by-step basis. This temporal discretization significantly reduces the computational complexity traditionally associated with solving both forward and backward time-dependent OCP problems.

By linearizing the convection term and employing a one-shot method for the linear optimal system, we achieve solutions at each time step efficiently. The effectiveness of our approach is demonstrated through a comprehensive analysis of velocity, volume flow, pressure plots, and WSS, which are presented with detailed interpretations. Our results show that the proposed method not only simplifies the computational process but also yields highly accurate solutions, making it a viable alternative to conventional techniques.

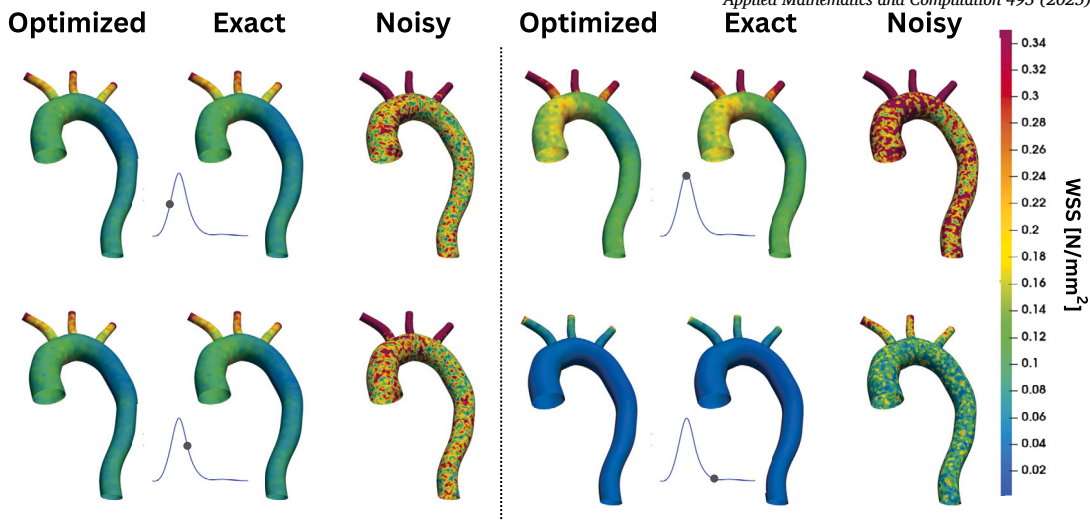

**Fig. 11.** Wall shear stress for $c = 1$ showing the reliability of the proposed scheme.

The results show that the optimization process consistently aligns the computed data with the exact data across various slices and SNR values. Specifically, for volume flow rates measured at different slices ($S_1$, $S_2$, $S_3$, and $S_4$), the optimized data closely followed the exact data, particularly at higher SNR values. This alignment is evident in the reduced variability and smoother profiles of the optimal data compared to the noisy data, indicating successful noise attenuation. For example, at an average SNR of 2.71, the optimal data exhibited minimal fluctuations and closely tracked the exact data across all slices, demonstrating the robustness of our optimization method.

Furthermore, the analysis of pressure data revealed a remarkable correspondence between the computed and exact pressure values. Utilizing 4D MRI-type synthetic data, we successfully extracted pressure values that closely matched the theoretical data, highlighting the potential of this methodology for accurate pressure estimation. This is crucial for solving optimal control problems in cardiovascular studies, where direct pressure measurements are often unavailable. The close match between the recovered and exact pressures validates our approach and underscores its applicability in clinical scenarios.

The velocity field analysis also demonstrated significant improvements through optimization. The optimized velocity magnitudes showed reduced discrepancies compared to the noisy data, aligning more closely with the exact solutions. This reduction in error margins reinforces the effectiveness of the optimization process in enhancing the accuracy of velocity measurements, which is critical for precise flow measurements and control in medical diagnostics and surgical planning.

In summary, our study advances computational techniques for cardiovascular modeling, specifically through an optimization approach that effectively reduces noise and improves the accuracy of volume flow rate, pressure, WSS and velocity measurements in the aorta. While the current application focuses on the cardiovascular system, the proposed method's potential extends to other domains where accurate flow simulation is critical, such as pulmonary, cerebrospinal, and even certain industrial fluid dynamics applications. These findings have significant implications for enhancing diagnostic and therapeutic strategies in clinical practice by providing a robust framework that balances computational efficiency with solution accuracy. Future work will focus on refining this approach further and applying it to more complex physiological scenarios using real 4D MRI data, ultimately aiming to improve patient outcomes in cardiovascular health and other domains where flow dynamics play a vital role.

### Acknowledgements

This work was supported by FCT – Fundacao para a Ciencia e Tecnologia through the CEMAT project PTDC/MAT-APL/7076/2020, https://doi.org/10.54499/PTDC/MAT-APL/7076/2020, and UIDB/04621/2020/IST-ID, https://doi.org/10.54499/UIDB/04621/2020.

### Data availability

Data will be made available on request.

### References

[1] P.R. Amestoy, I.S. Duff, J.Y. L'Excellent, J. Koster, A fully asynchronous multifrontal solver using distributed dynamic scheduling, SIAM J. Matrix Anal. Appl. 23 (2001) 15–41.

[2] A.E. Anderson, B.J. Ellis, J.A. Weiss, Verification, validation and sensitivity studies in computational biomechanics, Comput. Methods Biomech. Biomed. Eng. 10 (2007) 171–184.
[3] T.J. Barth, S. Mishra, C. Schwab, Uncertainty Quantification in Computational Fluid Dynamics, 2013.
[4] J. Bock, A. Frydrychowicz, A.F. Stalder, T.A. Bley, H. Burkhardt, J. Hennig, M. Markl, 4d phase contrast mri at 3 t: effect of standard and blood-pool contrast agents on snr, pc-mra, and blood flow visualization. Magnetic resonance in medicine: an official, J. Int. Soc. Magn. Reson. Med. 63 (2010) 330–338.
[5] B. Casas, J. Lantz, F. Viola, G. Cedersund, A.F. Bolger, C.J. Carlhäll, M. Karlsson, T. Ebbers, Bridging the gap between measurements and modelling: a cardiovascular functional avatar, Sci. Rep. 7 (2017) 1–15.
[6] S. De, F. Guilak, M.R. Mofrad, Computational Modeling in Biomechanics, Springer, 2010.
[7] P. Dyverfeldt, M. Bissell, A.J. Barker, A.F. Bolger, C.J. Carlhäll, T. Ebbers, C.J. Francios, A. Frydrychowicz, J. Geiger, D. Giese, et al., 4d flow cardiovascular magnetic resonance consensus statement, J. Cardiovasc. Magn. Reson. 17 (2015) 72.
[8] M. D'Elia, M. Perego, A. Veneziani, A variational data assimilation procedure for the incompressible Navier-Stokes equations in hemodynamics, J. Sci. Comput. 52 (2012) 340–359.
[9] E. Fevola, F. Ballarin, L. Jiménez-Juan, S. Fremes, S. Grivet-Talocia, G. Rozza, P. Triverio, An optimal control approach to determine resistance-type boundary conditions from in-vivo data for cardiovascular simulations, Int. J. Numer. Methods Biomed. Eng. 37 (2021) e3516.
[10] S.W. Funke, M. Nordaas, Ø. Evju, M.S. Alnæs, K.A. Mardal, Variational data assimilation for transient blood flow simulations: cerebral aneurysms as an illustrative example, Int. J. Numer. Methods Biomed. Eng. 35 (2019) e3152.
[11] T. Guerra, C. Catarino, T. Mestre, S. Santos, J. Tiago, A. Sequeira, A data assimilation approach for non-Newtonian blood flow simulations in 3d geometries, Appl. Math. Comput. 321 (2018) 176–194.
[12] T. Guerra, J. Tiago, A. Sequeira, Optimal control in blood flow simulations, Int. J. Non-Linear Mech. 64 (2014) 57–69.
[13] M. Gunzburger, Adjoint equation-based methods for control problems in incompressible, viscous flows, Flow Turbul. Combust. 65 (2000) 249–272.
[14] N.H. de Hoon, A.C. Jalba, E. Farag, P. van Ooij, A.J. Nederveen, E. Eisemann, A. Vilanova, Data assimilation for full 4d pc-mri measurements: physics-based denoising and interpolation, in: Computer Graphics Forum, Wiley Online Library, 2020, pp. 496–512.
[15] F.C. Hoppensteadt, C.S. Peskin, Mathematics in Medicine and the Life Sciences, vol. 10, Springer Science & Business Media, 2013.
[16] K. Itatani, S. Miyazaki, T. Furusawa, S. Numata, S. Yamazaki, K. Morimoto, R. Makino, H. Morichi, T. Nishino, H. Yaku, New imaging tools in cardiovascular medicine: computational fluid dynamics and 4d flow mri, Gen. Thorac. Cardiovasc. Surg. 65 (2017) 611–621.
[17] A. Manzoni, A. Quarteroni, S. Salsa, Optimal Control of Partial Differential Equations, Springer, 2021.
[18] M. Markl, S. Schnell, C. Wu, E. Bollache, K. Jarvis, A. Barker, J. Robinson, C. Rigsby, Advanced flow mri: emerging techniques and applications, Clin. Radiol. 71 (2016) 779–795.
[19] T. Otani, H. Yamashita, K. Iwata, S.Y. Ilik, S. Yamada, Y. Watanabe, S. Wada, A concept on velocity estimation from magnetic resonance velocity images based on variational optimal boundary control, J. Biomech. Sci. Eng. 17 (2022) 22–00050.
[20] M. Perego, A. Veneziani, C. Vergara, A variational approach for estimating the compliance of the cardiovascular tissue: an inverse fluid-structure interaction problem, SIAM J. Sci. Comput. 33 (2011) 1181–1211.
[21] A. Quarteroni, A. Valli, Numerical Approximation of Partial Differential Equations, vol. 23, Springer Science & Business Media, 2008.
[22] M.S. Rahman, M.A. Haque, Mathematical modeling of blood flow, in: 2012 International Conference on Informatics, Electronics & Vision, ICIEV, IEEE, 2012, pp. 672–676.
[23] P. Reymond, P. Crosetto, S. Deparis, A. Quarteroni, N. Stergiopulos, Physiological simulation of blood flow in the aorta: comparison of hemodynamic indices as predicted by 3-d fsi, 3-d rigid wall and 1-d models, Med. Eng. Phys. 35 (2013) 784–791.
[24] A. Sarrami-Foroushani, M.N. Esfahany, A.N. Moghaddam, H.S. Rad, K. Firouznia, M. Shakiba, H. Ghanaati, I.D. Wilkinson, A.F. Frangi, Velocity measurement in carotid artery: quantitative comparison of time-resolved 3d phase-contrast mri and image-based computational fluid dynamics, Iran. J. Radiol. 12 (2015).
[25] Z. Stankovic, B.D. Allen, J. Garcia, K.B. Jarvis, M. Markl, 4d flow imaging with mri, Cardiovasc. Diagn. Ther. 4 (2014) 173.
[26] J. Tiago, A. Gambaruto, A. Sequeira, Patient-specific blood flow simulations: setting Dirichlet boundary conditions for minimal error with respect to measured data, Math. Model. Nat. Phenom. 9 (2014) 98–116.
[27] J. Töger, M.J. Zahr, N. Aristokleous, K. Markenroth Bloch, M. Carlsson, P.O. Persson, Blood flow imaging by optimal matching of computational fluid dynamics to 4d-flow data, Magn. Reson. Med. 84 (2020) 2231–2245.
[28] F. Wang, Z. Dong, T.G. Reese, B. Bilgic, M. Katherine Manhard, J. Chen, J.R. Polimeni, L.L. Wald, K. Setsompop, Echo planar time-resolved imaging (epti), Magn. Reson. Med. 81 (2019) 3599–3615.